\documentclass{emulateapj}

\usepackage{psfig}

\newcommand{\etal}{\rm et~al.\ } 
\newcommand{\blc}{\ensuremath{B_{\rm LC}}}
\newcommand{\eav}{\ensuremath{\langle E \rangle}}

\newcommand{\dmunit}{\,pc\,cm$^{-3}$}
\newcommand{\jyms}{Jy$\rm \mu$s}
\newcommand{\gree}{$\gamma$-ray}

\newcommand{\psrc}{PSR~J1824$-$2452A}

\slugcomment{To be submitted to \apj}

\shorttitle{Giant Pulses from \psrc}
\shortauthors{Knight \etal}

\begin{document}

\title{A Study of Giant Pulses from \psrc}

\author{H. S. Knight\footnote{Affiliated with the Australia Telescope
National Facility, CSIRO}\footnote{ Email: hknight@astro.swin.edu.au}, M. Bailes}
\affil{Centre for Astrophysics and Supercomputing, Swinburne University of Technology,
P.O. Box 218, Hawthorn VIC 3122, Australia}
\author{R. N. Manchester}
\affil{Australia Telescope National Facility, CSIRO, P.O. Box 76, Epping NSW 1710, Australia}
\and
\author{S. M. Ord}
\affil{School of Physics, University of Sydney, NSW 2006, Australia}

\begin{abstract}
We have searched for microsecond bursts of emission from millisecond
pulsars in the globular cluster M28 using the Parkes radio telescope.
We detected a total of 27 giant pulses from the known emitter \psrc.
At wavelengths around 20\,cm the giant pulses are scatter-broadened to
widths of around 2\,$\mu$s and follow power-law statistics.  The
pulses occur in two narrow phase-windows which correlate in phase with
X-ray emission and trail the peaks of the integrated radio
pulse-components.  Notably, the integrated radio emission at these
phase windows has a steeper spectral index than other emission.  The
giant pulses exhibit a high degree of polarization, with many being
100\% elliptically polarized.  Their position angles appear random.
Although the integrated emission of \psrc\ is relatively stable for
the frequencies and bandwidths observed, the intensities of individual
giant pulses vary considerably across our bands.  Two pulses were
detected at both 2700 and 3500\,MHz.  The narrower of the two pulses
is 20\,ns wide at 3500\,MHz.  At 2700\,MHz this pulse has an inferred
brightness temperature at maximum of $5 \times 10^{37}$\,K.  Our
observations suggest the giant pulses of \psrc\ are generated in the
same part of the magnetosphere as X-ray emission through a different
emission process to that of ordinary pulses.
\end{abstract}


\section{Introduction}\label{sec:introduction}

The Crab pulsar was discovered through the observation of a series of
strong bursts of radio emission \citep{sr68}.  The sampling interval
of this search was about 18 times the 33\,ms pulse period, so if all
of its pulses had been equal to the average pulse it would have been
rendered invisible.  Instead, the pulses followed power-law
statistics, and so some pulses with energies many times stronger than
the average pulse-energy were detected.  These giant pulses (GPs) are
composed of $\lesssim 2$\,ns ultra-bright shots \citep{hkwe03} that
can occur in emission envelopes that are as long as $\sim 100$\,$\mu$s
\citep{sbh+99,kni06}.  Propagative effects have a frequency-dependent
nature.  As the spacing and number of shots are frequency independent,
the observed multiplicity of components is due to the emission
mechanism itself \citep{sbh+99}.

The GPs are most readily found in the main-pulse and inter-pulse radio
components, which correlate in phase with the optical, X-ray, and
\gree\ pulse peaks \citep{lcu+95}.  Other radio pulse-components do
not have associated high-energy emission, but some GPs are seen at
their phases \citep{jsk+05}.

The individual nano-shots of GPs from the Crab pulsar have a high
degree of circular polarization and come in both orientations
\citep{hkwe03}.  The distribution in degree of circular polarization
of unresolved GPs can be accurately modelled by assuming
superpositions of many ($\sim 100$) nano-pulses that are 100\%
circularly polarized \citep{psk+06}.

The millisecond pulsars B1937+21, J1824$-$2452A, J1823$-$3021A,
J0218+4232, and B1957+20 have also been shown to emit short-timescale
pulses with inferred brightness-temperatures ($T_{B}$) that are very
high \citep{cstt96,rj01,kbmo05,kbm+06,kni06}.  Although many of these
ultra-bright pulses are weaker in pulse-energy terms than the average
pulse (see, e.g. Soglasnov \etal 2004)\nocite{spb+04}, they can be
defined as GPs through a number of other criteria.  These are that
they are always unresolved on microsecond timescales, occur in narrow
phase-windows that align with X-ray emission, and that their energies
follow power-law statistics.  Unlike the GPs of the Crab pulsar, the
GPs from the millisecond pulsars do not occur at all phases where
integrated emission is seen, and do not contribute markedly to the
integrated emission profile.  For PSR~B1937+21 the GPs occur at the
extreme trailing edges of the two integrated emission components.
\citet{spb+04} found that the GPs from PSR~B1937+21 usually consist of
single $\sim 16$\,ns shots, but also presented a pulse with structure
over hundreds of nanoseconds.  \citet{pskk04} found that at 600\,MHz
the GPs from PSR~B1937+21 are $< 10$\% linearly polarized and highly
circularly polarized.

\psrc\ is an isolated millisecond pulsar with a period of
$P=3.054$\,ms.  It is located 5.6\,kpc from the Sun in the globular
cluster M28 \citep{lbm+87,har96a}.  Acceleration in the potential of
its host cluster means the period derivative ($\dot{P}$) contribution
from spin down is not directly observable, and consequently other
derived parameters are not known with certainty.  However, indirect
evidence exists that the large $\dot{P}$ observed is indicative of the
true value --- \psrc\ glitches \citep{cb04}, emits GPs, emits strong
X-ray pulses \citep{rjm+98}, and has a large second period derivative
($\ddot{P}$) \citep{cbl+96}.  If the observed $\dot{P}$ is similar to
the intrinsic value, then \psrc\ is around 30\,Myr old and it and
PSR~J1823$-$3021A are the youngest millisecond pulsars known.  This
assumption also means that PSRs J1824$-$2452A and B1937+21 have
inferred magnetic fields at their light cylinders ($B_{\rm LC} \propto
P^{-2.5}\dot{P}^{0.5}$) and spin-down luminosities ($\dot{E} \propto
P^{-3}\dot{P}$) that are the highest of all millisecond pulsars.

\citet{rj01} (hereafter RJ01) discovered GPs at a center frequency
($\nu$) of 1520\,MHz from \psrc\ with energies as low as 30 times the
mean pulse-energy, \eav.  Use of a hardware filter-bank on a high
dispersion measure source ($\sim 120$\dmunit) limited their effective
time resolution to 200\,$\mu$s and they were unable to resolve the
pulses.  Six of the eight brightest pulse detections fell within a
single phase-window, which slightly trailed one of the three pulse
components.  To date, it has been unclear how narrow the GPs from
\psrc\ are, and whether they have substructure like GPs from the Crab
pulsar.

In this paper we use a baseband recorder to achieve much finer time
resolution than RJ01 in an attempt to uncover and study a larger
population of GPs emanating from millisecond pulsars in M28.  We
resolve the GPs from \psrc, better quantify their phase and energy
distributions, and perform a polarimetric analysis.

\section{Observations and Data Analysis}\label{sec:observations}

Observations were taken using the Parkes 64-m radio telescope between
2003 February and 2004 December.  All data were taken using the
Caltech-Parkes-Swinburne-Recorder Mark II (CPSR2) \citep{bai03}.  This
is a baseband recorder that two-bit samples one or two
dual-polarization 64\,MHz bands at the Nyquist rate.  Coherent
dedispersion \citep{hr75,sta98,van03a} can be performed on the input
stream to remove the cold-plasma dispersion of the interstellar medium
(ISM) to give an effective sampling rate of 7.8125\,ns per 64\,MHz
band.  However, it is more advantageous with respect to interference
removal and computing requirements to form a ``coherent filter-bank''
that gives enhanced frequency resolution at the cost of reduced time
resolution.  In the initial searches a 256-channel filter bank was
formed at a DM of 119.868\dmunit\ and searched for bursts of emission
at sampling times of 4 -- 128\,$\mu$s.  Initial searches were carried
out over a DM range of 0.1\dmunit, but later searches used increased
ranges of up to 6.0\dmunit.  Within these ranges, coherent
dedispersion and channel-summing were repeatedly performed so as to
ensure dispersive smearing across the entire band was always less then
4\,$\mu$s.  PSR~J1824$-$2452F, which has a DM of 123.8\dmunit\ (Ransom
\etal 2005), is an example of a pulsar whose DM meant that these
searches probably did not probe its emission effectively.  Data were
searched with a variety of algorithms, and in general data files
without candidates were deleted.  Subsequently, the remaining data
files were re-dedispersed multiple times and searched at time
resolutions down to 0.5\,$\mu$s so as to remove dispersive smearing
for DMs between 116.9 and 124.9\dmunit.  Table~\ref{tab:tsys}
summarizes the data taken.  Columns 1 and 2 respectively show the
observation date and receiver used.  All receivers used have two
orthogonal linear probes.  The center frequency observed and system
equivalent flux density (SEFD) for the on-source observations are
shown in columns 3 and 4 respectively.  Columns 5 and 6 show the total
duration of the initial and secondary searches respectively, and
column 7 shows the flux density of \psrc.  The criteria for a
detection leading to human scrutiny in the secondary search was two
consecutive samples adding to give a spike 13 times the RMS.  The
detection threshold for 4\,$\mu$s sampling is shown in column 8 in
absolute terms, and in column 9 in terms of the mean pulse-energy of
\psrc.  The radiometer equation (see, e.g. Edwards 2001)\nocite{edw01}
dictates that the minimum energy detectable varies as $t_{\rm
samp}^{0.5}$, where $t_{\rm samp}$ is the sampling time used.
Therefore the threshold for the later searches, which had shorter
sampling times, was lower for unresolved pulses.

\section{Search Results}\label{sec:results}

At 1341 -- 1720\,MHz (hereafter 20\,cm) 25 pulses were detected and
confirmed to be associated with \psrc\ by virtue of occurring at its
DM.  Of these, 21 were seen in 2003 and 4 were seen in 2004.  In
addition, two pulses from \psrc\ were observed at 2700\,MHz.
Counterparts to these two pulses were revealed in 3500\,MHz data upon
further inspection.  No pulses were found from other sources in the
cluster.  PSR~J1824$-$2452C is $\sim 170$\,$\mu$Jy at 1950\,MHz and is
one of the brightest of the other pulsars (Ransom \etal 2005; Ransom
2006, priv. comm.)\nocite{rhs+05b}.  If a spectral index ($\alpha_{\rm
spec}$) of $-1.9$ is assumed for this 4.159\,ms pulsar, its average
flux-density at 1341\,MHz is 350\,$\mu$Jy and its average pulse energy
is 1.4\,\jyms.  Therefore, our 91\,\jyms\ threshold in
Table~\ref{tab:tsys} at 1341\,MHz corresponds to a threshold of
60$\eav$.  Although this is not particularly constraining given the
paucity of such strong pulses emitted by the known GP emitters PSRs
J0218+4232 and B1957+20 \citep{kbm+06}, it can be said that
PSR~J1824$-$2452C is unlikely to emit GPs at high rate.

\subsection{Phases}\label{sec:phases}

\begin{figure}
  \begin{center}
    \includegraphics[scale=0.6,angle=270]{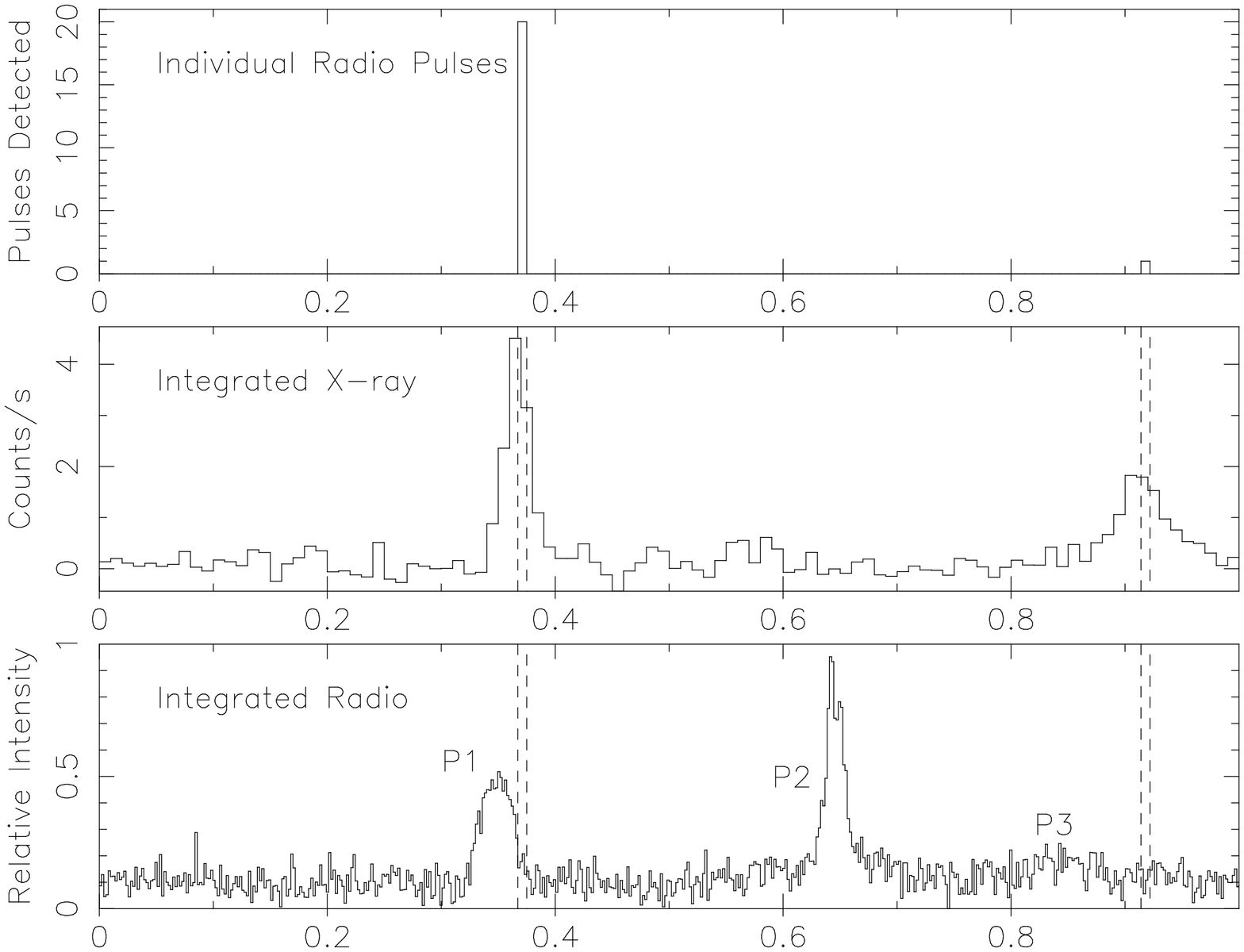}
    \caption{ Relative phases of emission from \psrc.  (a) Phases of
    giant pulses detected in 2003 February and 2003 April.  (b) The
    RXTE 2 -- 16\,keV pulse profile of \psrc\ reproduced from
    \citet{skr+01}.  Phase alignment was performed by noting that in
    the data of \citet{rjm+98} the peak of the X-ray profile trails
    the leading radio-pulse by 0.017 pulse periods.  The X-ray phases
    are accurate to around 10\,$\mu$s.  Rots \etal determined this
    alignment using the radio ephemerides of \citet{bs97}.  (c)
    Mean pulse profile at 1341\,MHz created from 2003 April data.}
    \label{fig:xray} \end{center}
\end{figure}

The phases of the pulses detected from \psrc\ in 2003 February and
2003 April are shown in Figure~\ref{fig:xray}.  We follow the
numbering convention of \citet{bs97} and denote the leading
steep-spectrum component at phase 0.35 as P1, and the trailing two
components as P2 and P3 respectively.  The pulse population at phase
0.37 occurs over 0.006 periods (18\,$\mu$s) and trails the center of
the P1 component by around 0.02 periods.  Both of the pulses seen at
2700\,MHz also come from the vicinity of this window.  Despite these
two pulse detections at 2700\,MHz, phase-coherent summation of
$3\times 10^{6}$ pulses was not enough to provide a detection of the
accompanying P1 component.  If \psrc\ did not emit P2 pulses, at high
frequency it would be more readily detected through individual GPs
than through its integrated emission.

The main-pulse GPs of PSR~B1937+21 occur over a window of similar
width to that of \psrc\ --- 0.007 periods \citep{spb+04}.  The GP
emission zones of \psrc\ have nearly equal phases to the peaks of the
X-ray pulses.  We consider that any misalignment is close to the error
in the phase determination.  The 1341\,MHz pulse detected at phase
0.92 was $21\eav$ and well above the detection threshold.  Another
pulse with a similar phase was detected in 2004 November, and RJ01
also detected a pulse here.  Therefore, there is definitely a
secondary GP population around phase 0.92.

\subsection{Energies}\label{sec:energies}

At a sampling interval of 12\,$\mu$s all visible 20\,cm pulses had
intensities above 50\% of the peak intensity for just one or two
samples.  This sampling interval was therefore used to determine the
FWHM energies of these pulses.  The most energetic pulse was $91\eav$
and the weakest was $8.4\eav$.  At high frequency our estimates are
uncertain due to the poor signal to noise ratio of the integrated
profile, but the two pulses seen at 2700\,MHz (hereafter pulses 1 and
2) had energies around 260 and 114\,$\eav$ respectively.  Their
counterparts at 3500\,MHz were discovered by analysing the profiles of
the pulses the GPs occurred in, and were 104 and 105\,$\eav$
respectively.

At $\nu \ge 2700$\,MHz only one component is visible in the integrated
profile of \psrc, and this is taken to be the P2 component.  No GPs
occur at the phase of this component, and so its average energy may
not be related to the GP energies.  The average P1 energy, $\langle
E_{\rm P1} \rangle$, may be a better measure of how strong a GP is.
At 1341\,MHz 36\% and 49\% of the total radio flux-density ($S$) of
\psrc\ is emitted in the P1 and P2 components respectively.
Extrapolation can be performed using the $\alpha_{\rm spec\ (P1)} =
-3.1$ and $\alpha_{\rm spec\ (P2)} = -1.4$ spectral indices found by
\citet{ffb91}.  At 2700\,MHz $S_{P2}/S_{P1}$ is therefore larger than
at 1341\,MHz by a factor of around $(2700/1341)^{3.1-1.4} = 3.3$.  The
$260\langle E_{\rm P2} \rangle$ pulse seen at 2700\,MHz is equivalent
to a $1200\langle E_{\rm P1} \rangle$ pulse at 1341\,MHz.  This is a
$\gtrsim 400 \eav$ pulse, which is far stronger than any of the pulses
seen at 20\,cm.  Therefore, the GP must be stronger relative to the
average pulse at high frequency than at low frequency.  This means
that GPs are probably strong only over moderate bandwidths.  Further
evidence supporting this hypothesis is the fact that the GP had a
steep spectral index of $-5.4$.

\subsection{Rates}\label{sec:rates}

A probability distribution for pulse energies following power-law
statistics can be expressed in the form:

\begin{equation}
P(E > E_{0}) = K E_{0}^{-\alpha},
\end{equation}

\noindent where $E_{0}$ is measured in terms of the mean pulse energy.
Integration gives the fraction of radio energy emitted
in the form of GPs of energies greater than $E_{0}$,

\begin{equation}
S_{\rm GP}(E > E_{0}) = \frac{K \alpha}{\alpha-1} E_{0}^{1-\alpha}.
\end{equation}

\begin{figure}
  \begin{center} \includegraphics[scale=0.6,angle=270]{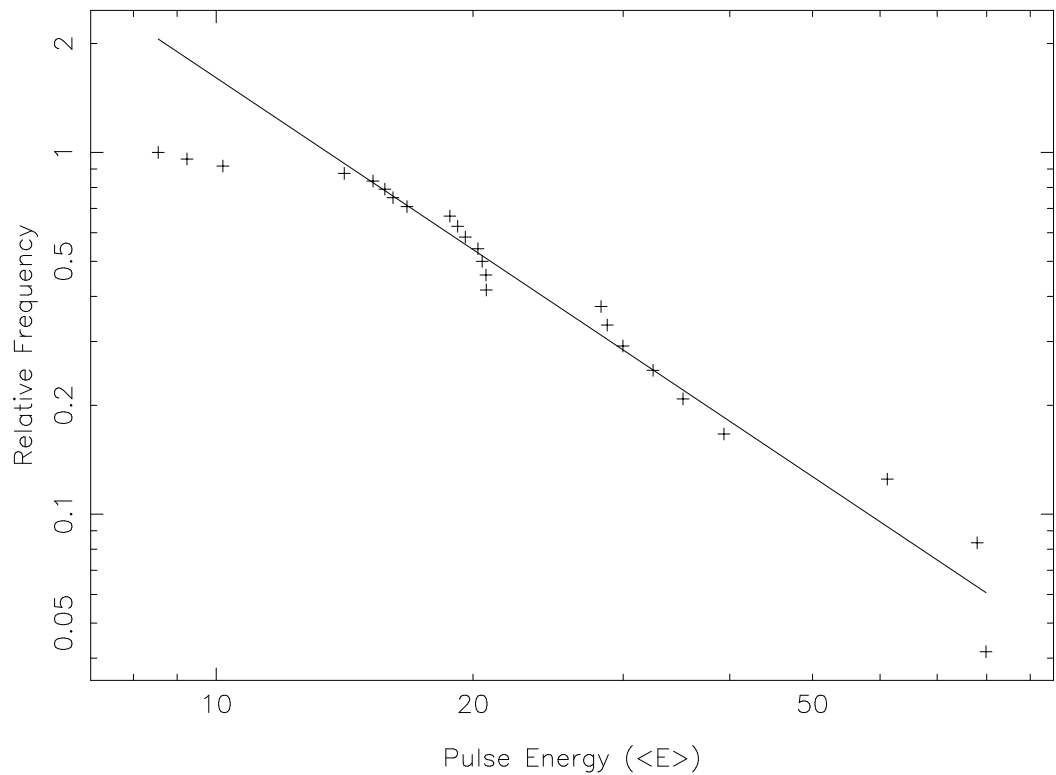}
    \caption{Cumulative probability distribution of 20\,cm pulse
    energies.  The solid line denotes the line of best fit, excluding
    the 3 pulses with the lowest energies, and the 3 pulses with the
    highest energies.  It has a slope of $-1.6$.  } \label{fig:cumu}
    \end{center}
\end{figure}

\noindent A cumulative probability distribution for the 20\,cm pulses
is shown in Figure~\ref{fig:cumu}.  Exclusion of the 3 most energetic
and 3 least energetic pulses yields a best-fit power-law with $K =
2.5\times 10^{-4}$ and $\alpha = 1.6$.  Both the plot and the energy
thresholds of Table~\ref{tab:tsys} indicate pulses below 22\,$\eav$
have been missed.  Exclusion of the 8 least energetic pulses gives a
similar fit but exclusion of 15 low-energy pulses gives an alternative
fit with $\alpha = 2.3$.  Power-law indices in the range 1.4 -- 2.5
have been fitted for pulsars that emit GPs \citep{kbm+06}, and so both
fits are plausible.  By comparison RJ01 suggested $\alpha$ is in the
range of 3 -- 5.  This range of values is inconsistent with our data.

Assuming the original $\alpha = 1.6$ fit, $P(E>28\eav) = 1.3\times
10^{-6}$, which is similar to the value of $\sim 8.5 \times 10^{-7}$
of RJ01.  We find that \psrc\ emits GPs with total energies of
$20\eav$ at about 0.5 and 3 times the rates of PSRs B1937+21 and
J0218+4232 respectively \citep{spb+04,kbmo05}.

The overall contribution of GPs to the pulse profile of \psrc\ is
difficult to ascertain because they lie in a region where ordinary
emission occurs.  Analysis of the pulse profile suggests that at
1341\,MHz the integrated flux-density averages $\sim 5$\,mJy over the
GP phases.  The flux-density integrated over this window makes up
$\sim 3$\% of the flux-density integrated over the entire profile.
For GP emission contributing 100\%, 10\%, and 1\% of the total
flux-density over the GP phases, equation (2) gives low-energy
cut-offs of $1.7\times 10^{-3}$, $8.8\times 10^{-2}$, and 4.6\,$\eav$
respectively.  \citet{spb+04} found that the GP distribution of
PSR~B1937+21 probably extends to energies of 0.016 -- 0.032\,$\eav$.
If a similar cut-off occurs for \psrc\ then a 20 -- 30\% increase in
flux density would be expected at the phase of the GPs.  Such a bump
may be detectable by 100\,m class telescopes.

\section{Pulse Widths}\label{sec:widths}

\begin{figure}
  \begin{center} \includegraphics[scale=0.6,angle=270]{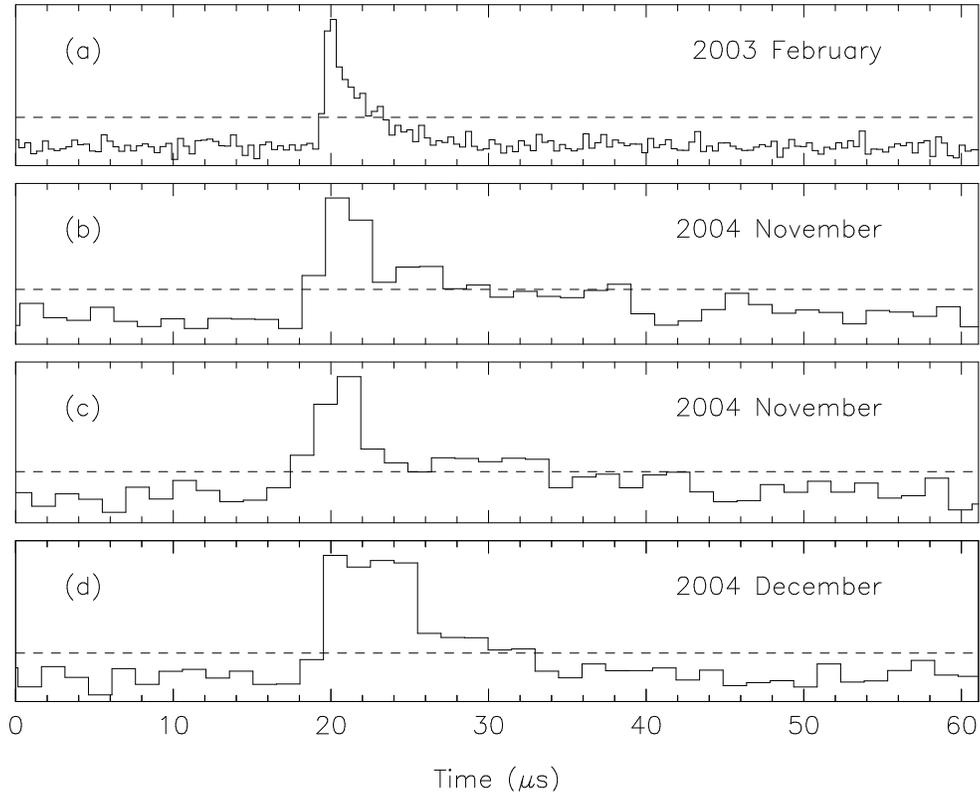}
    \caption{Variable scattering timescales between 2003 and 2004.
(a): Pulse observed in 2003 February at 1341\,MHz.  (b) and (c):
Pulses observed in 2004 November at 1341\,MHz.  (d): Pulse observed in
2004 December at 1405\,MHz. The dashed line indicates 30\% of the peak
intensity. }
    \label{fig:widths}
  \end{center}
\end{figure}

Pulse profiles binned at 750\,ns gave full widths at half-maximum
(FWHM) for the 20\,cm pulses of between 750\,ns and 6.0\,$\mu$s.  The
FWHM of the strongest 2003 pulses were 2.2 and 1.1\,$\mu$s at 1341 and
1720\,MHz respectively.  Although these widths imply a flatter
frequency dependence than the $\nu^{-4.4}$ spectrum of
Kolmogorov-spectrum scattering \citep{ric96}, we think that more
pulses need to be studied to make definite conclusions.  A 1341\,MHz
pulse representative of other 2003 pulses with regard to typical width
and shape is shown in Figure~\ref{fig:widths}, along with several
pulses observed in 2004.  Emission above 30\% of the peak is seen at
long delays for all 2004 pulses but not for any 2003 pulses.  The
exponential decay time of the scattering tail therefore substantially
changes on $\sim 1$\,yr timescales.

\begin{figure}
  \begin{center} \includegraphics[scale=0.6,angle=270]{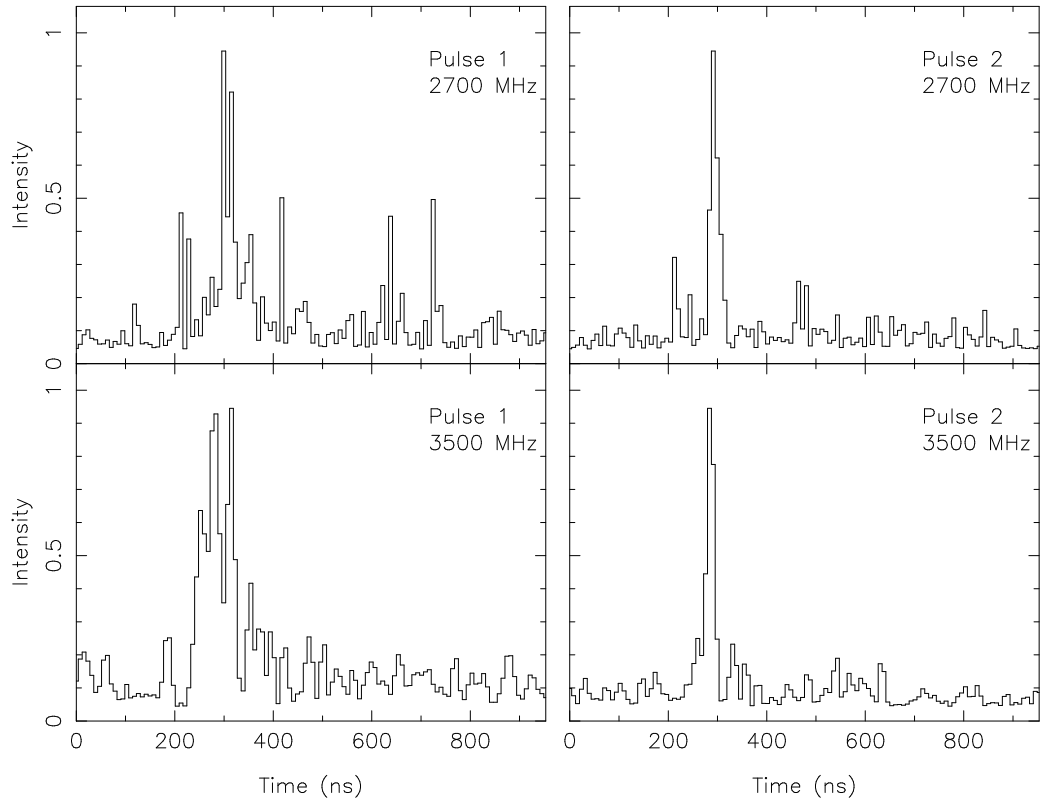}
    \caption{Pulses observed at 2700 and 3500\,MHz in 2003 December shown
at a time resolution of 7.8125\,ns.}
    \label{fig:quadpulse}
  \end{center}
\end{figure}

The high-frequency observations taken in 2003 December used two bands
separated by 800\,MHz.  By aligning pulses from the two bands we
obtained a DM of $119.8688 \pm 0.0002$ for this epoch, sufficiently
accurate to allow dedispersion to remove all dispersive effects of a
cold ISM to the 7.8125\,ns Nyquist sampling-rate for each band.  Fully
dedispersed profiles of \psrc\ are shown in
Figure~\ref{fig:quadpulse}.  Pulse 1 appears to be intrinsically
broader than pulse 2 and at 2700\,MHz has more unresolved structure
occurring hundreds of nanoseconds after the main pulse.  A
$\nu^{-4.4}$ extrapolation of the 1.1\,$\mu$s FWHM at 1720\,MHz
predicts widths of 150 and 48\,ns at 2700 and 3500\,MHz respectively.
Given that the structure disappears at 3500\,MHz we consider that it
is probably a manifestation of scattering in the ISM.  Alternatively
it could be narrow-band, or just receiver noise.

At high frequencies the pulse width does not seem to scale with
frequency --- pulse 1 at 3500\,MHz is broader than either pulse at
2700\,MHz.  We may therefore be resolving the pulse envelopes.  If
this is true, moderately enhanced time resolution could reveal whether
the emission consists of short strong bursts interspersed by
intervals where there is little or no emission.  Our observations
provide no evidence that pulse 2 is anything other than a single
coherent shot.  It has a FWHM of 20\,ns at 3500\,MHz, which is similar
to the $\sim 15$\,ns inferred for the GPs of PSR~B1937+21
\citep{spb+04}.  The pulse reaches 2300\,Jy and therefore has an
inferred brightness temperature at maximum of $5 \times 10^{37}$\,K,
which is higher than the nano-pulses directly observed from the Crab
pulsar \citep{hkwe03}.  However, simulations of the ISM response of
GPs from PSR~B1937+21 by Soglasnov \etal indicate that its GPs can
reach $T_{B} \sim 5 \times 10^{39}$\,K, which is much brighter.

\section{Polarimetry}\label{sec:polarimetry}

\begin{figure}
  \begin{center} \includegraphics[scale=0.6,angle=270]{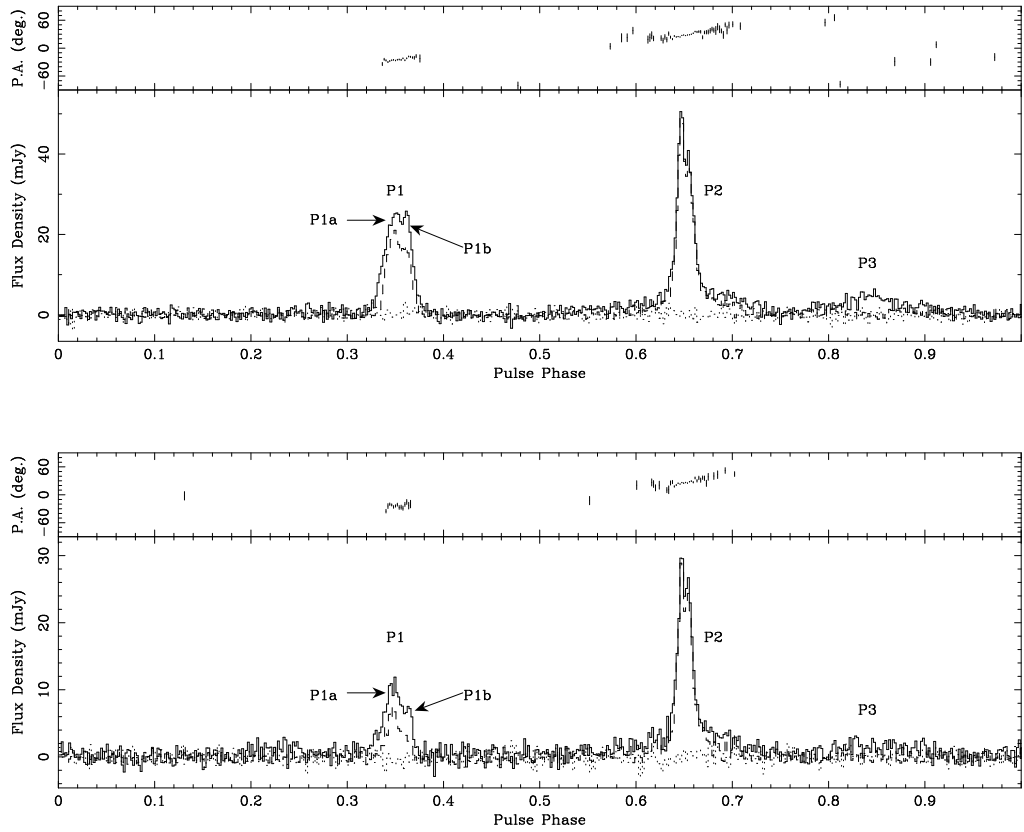}
    \caption{Polarization plots of 2003 April observations of \psrc\ at
    1341\,MHz (top) and 1720\,MHz (bottom).  The solid, dashed, and
    dotted lines denote total, linearly polarized, and circularly
    polarized emission respectively.  The position angle corrected for
    Faraday rotation is shown at the top of each profile.  The
    position angles shown are not absolute.}  \label{fig:manchester}
    \end{center}
\end{figure}

The 2003 April GPs and integrated profiles were calibrated using the
{\sc psrchive} software library \citep{hvm04} using a similar
methodology to \citet{ovhb04}.  The 1341 and 1720\,MHz polarimetric
profiles of \psrc\ are shown in Figure~\ref{fig:manchester}.  At
1341\,MHz the high flux-density portions of the P1 and P2 pulse
components are respectively 72\% and 96\% linearly polarized.  We
detect no circular polarization in any component and no polarization
at all in the P3 component.  The position angle swing between the
center of the P1 component and the peak of the P2 component is
48$^{\circ}$ at 1341\,MHz, which is similar to that presented by other
authors \citep{bs97,stc99,ovhb04}.  We find no evidence in our data
for the mode-changing described by \citet{bs97}.  The P1b component of
the 1720\,MHz profile is reduced in intensity when compared to the
1341\,MHz profile.  The peak flux-density in this phase zone scales as
$\nu^{-4.4}$, whereas the flux-densities of the earlier P1 (P1a) and
the P2 peaks scale as $\nu^{-2.8}$ and $\nu^{-1.9}$ respectively.
\citep{kt00} found that on average GPs from PSR~B1937+21 have spectral
indices steeper than the integrated emission.  However, this cannot be
the case for the GPs of \psrc\ --- a power-law steeper than
$\nu^{-4.4}$ would have precluded our detections of GPs above 2\,GHz.

\begin{figure}
  \begin{center} \includegraphics[scale=0.6,angle=270]{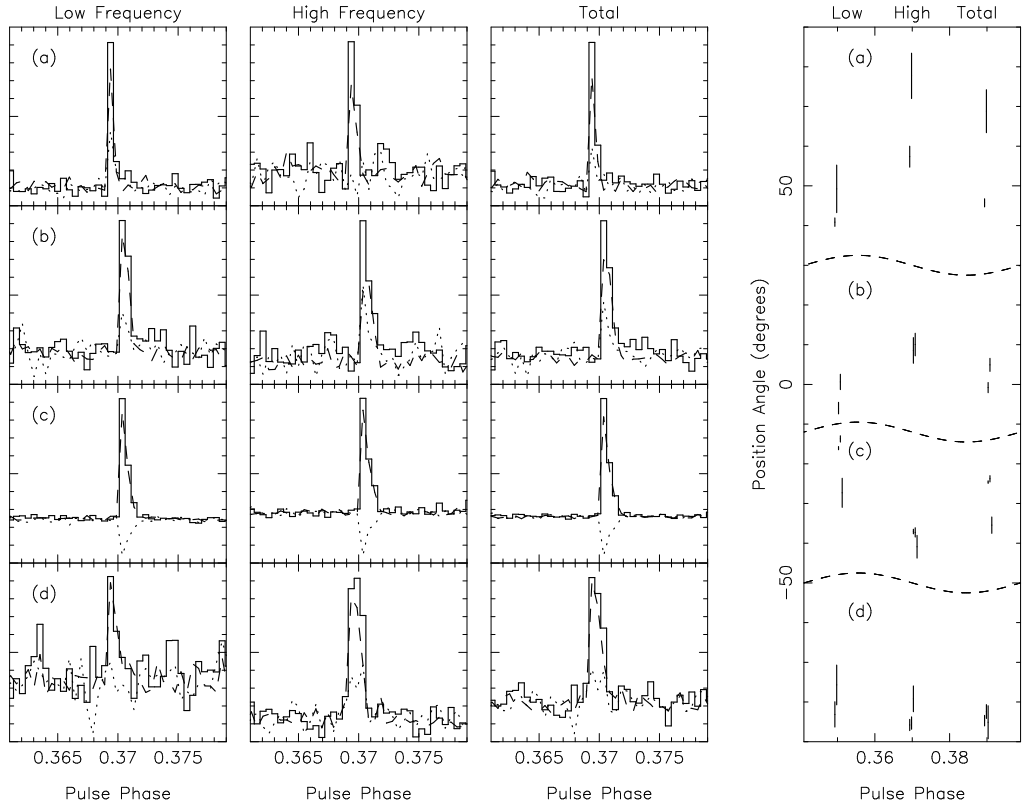}
    \caption{Polarization plots and position angles of four strong GPs
    observed at 20\,cm.  The left three columns show polarization
    plots of the GPs.  The first two columns respectively show the
    polarization of each GP in the low-frequency and high-frequency
    halves of the observing band.  The third column shows the
    polarization of each GP over the full band.  Total intensity,
    linear polarization, and circular polarization are shown by solid,
    dashed, and dotted lines respectively.  The fourth column shows
    the position angles of these pulses --- each of sections (a)-(d)
    contains only position angles of the corresponding GP.  For
    clarity the phases of the low-frequency and total position angles
    have been shifted by 0.02 pulse periods to earlier and later
    phases respectively.  The GPs clearly have different position
    angles to each other.  However, the position angle of each pulse
    is similar for each independent frequency band.  Noise cannot
    explain this consistency, so the position angles shown are
    intrinsic to the pulses.  The position angles are spread over all
    possible values, so it can be concluded that the emission
    mechanism of the GPs produces randomly distributed
    position angles.  } \label{fig:gppa} \end{center}
\end{figure}

By maximizing the linear polarization of the mean pulse-profile we fit
a rotation measure (RM) of $78.5 \pm 0.9$\,rad\,m$^{-2}$.  The RM of
$1 \pm 12$\,rad\,m$^{-2}$ published by \citet{rl94} is wrong.
Polarization properties of 13 GPs are summarized in
Table~\ref{tab:pol}.  The first three rows of data respectively show
the linear, circular, and total polarization fractions.  The last row
shows the ratio of circular to linear polarization.  Minimum, mean,
and maximum fractions for each of these quantities are shown in
columns 2, 3, and 4 respectively.  The GPs have both signs of circular
polarization.  \citet{hkwe03} and \citet{pskk04,psk+06} studied the
polarization of GPs of PSR~B1937+21 and the Crab pulsar, but neither
group reported significant linear polarization. By contrast, the GPs
of \psrc\ are highly elliptically polarized.  Some are even 100\%
linearly polarized.  The position angles of the GPs appear to be
random, as shown in Figure~\ref{fig:gppa}.  This could be due to a
propagation effect, or it could be intrinsic to the emission mechanism
of the GPs.  In the later case, it would mean that the preferred axis
of the emission region varies --- i.e. a strong local field
temporarily dominates over the global dipole.  This could be
indicative of small-scale plasma turbulence.

\section{Frequency Variability}\label{sec:variability}

\begin{figure}
  \begin{center}
    \includegraphics[scale=0.6,angle=270]{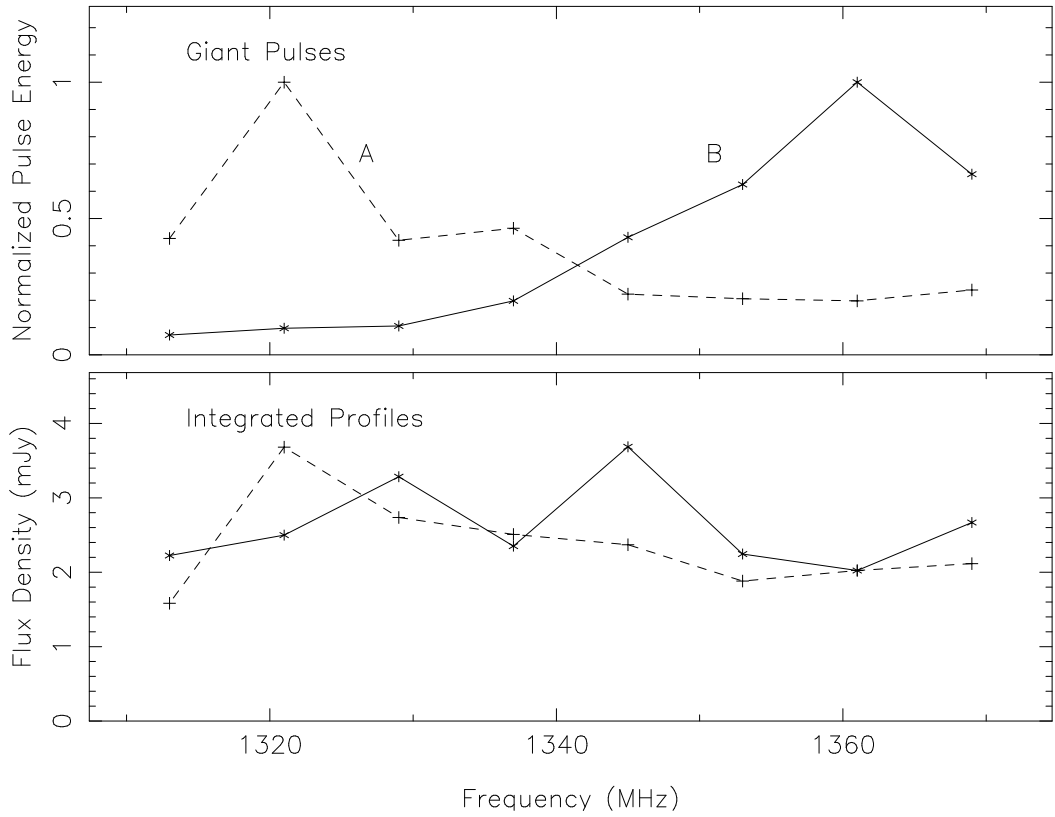}
    \caption{Frequency variability of two pulses observed in 2003
    April.  Top: Spectra of pulse A (dotted line and plus symbols) and
    pulse B (solid line and asterisks).  The energies for frequencies
    above 1340\,MHz (below 1330\,MHz) for pulse A (pulse B) are upper
    limits only.  Bottom: Integrated flux-densities for 300\,s of data
    about pulse A (dotted line and plus symbols) and pulse B (solid
    line and asterisks).}
    \label{fig:variability}
  \end{center}
\end{figure}

The time scale of diffractive scintillation for \psrc\ at $\sim
1400$\,MHz is $\Delta t_{d} \sim 100$\,s \citep{cbl+96}.  The
corresponding frequency scale is given by $\Delta \nu_{d} = 1/2\pi
\tau_{b}$ \citep{ric96}.  Here $\tau_{b} \sim 2.5$\,$\mu$s is the
scattering timescale, so $\Delta \nu_{d} \sim 0.1$\,MHz.  Both $\Delta
t_{d}$ and $\Delta \nu_{d}$ are much smaller than our observation
timescales and bandwidths.  Little variability is therefore expected
in our observations.  Breaking the 2003 April observation into 300\,s
sections confirmed that there was no variability beyond what can be
explained by a low signal to noise ratio.  In contrast, the GPs vary
widely in their spectral characteristics.  Frequently pulses observed
in one 64\,MHz band are not detected at all in the other 64\,MHz band.
Pulse intensities also tend to drop to below our detection
sensitivities within the bands.  The top panel of
Figure~\ref{fig:variability} shows two pulses observed 3900\,s apart
that have nearly inverted frequency dependencies.  Pulse B is at least
10 times stronger at 1361\,MHz than it is at 1313\,MHz.  The lower
panel shows that the integrated emission in the 300\,s window around
the pulse only varies by around a factor of $\sim 2$ across the band.
The variability seen in the upper panel therefore cannot be due to
propagation in the ISM or variability of all emission.  It is
restricted to the GPs and is therefore probably intrinsic to their
emission mechanism.

\section{Discussion}\label{sec:discussion}

Our simultaneous observations of GPs at 2700 and 3500\,MHz show that
the GPs are not a narrow-band phenomena.  However, the disappearance
of the pulses within our 64\,MHz bands at 20\,cm indicates that
detection at any frequency is almost entirely dictated by the
narrow-band ($\lesssim 64$\,MHz) frequency characteristics of the
pulses.  Perhaps then the GPs are composed of finely spaced
nano-pulses that have narrow bandwidths?  If each such nano-pulse
peaks at a different frequency, the GPs could vary in strength over
small bandwidths, but be detectable over large bandwidths.  A similar
phenomenom occurs for the Crab pulsar \citep{eahw02,jsk+05}.  Testing
this hypothesis requires observations of high time-resolution.  These
would need to be taken simultaneously over multiple frequency bands so
that detection of different nano-pulses at different frequencies could
be distinguished from DM errors.  Another consequence of narrow-band
nano-pulses would be a frequency-dependent position angle.  Detection
of such an effect would require simultaneous polarimetric observations
over multiple frequency bands.  Notably, none of the pulses we
observed gave a satisfactory RM fit, so our observations do not rule
out such a hypothesis.

Our observations at 3500\,MHz show GPs from \psrc\ that are at most
100\,ns wide.  The Crab pulsar spins just $\sim 10$ times slower than
\psrc, yet it emits in envelopes that are up to $\sim 1000$ times
broader!  This means that envelope width is unlikely to scale as a
linear function of period.

Rate of GP emissivity cannot even be moderately dependent on \blc.
This is because \psrc\ emits a 20$\eav$ pulses 4000 times less
frequently than the Crab pulsar, despite its \blc\ being 80\% that of
the Crab pulsar \citep{kbm+06}.  The Crab pulsar has a spin-down
luminosity $\sim 200$ times larger than \psrc, so although $\dot{E}$
seems a better determinant of GP emissivity than \blc, it alone cannot
account for the $\sim 4000$ factor in the emissivity rates.  We found
that the GPs from \psrc\ occur at pulse phases where the integrated
emission has the steepest spectra.  Perhaps then this is a factor?  In
the model of GP emission of \citet{pet04}, photons are Compton
scattered from low frequencies to high frequencies.  Pulsars with
steep spectra therefore have the largest capacities for amplification,
and so GPs should be seen most readily for steep-spectrum sources.
This is indeed what is seen --- the millisecond pulsars that emit GPs
all have steep spectra \citep{kbm+06}.  However, the correlation is by
no means compelling --- further detections of emitters are needed to
help determine if $\alpha_{\rm spec}$ really contributes to
emissivity.

All GP components of millisecond pulsars appear to have associated
X-ray components and vice-versa.  The simplest interpretation of this
is that GP emission originates in the same part of the magnetosphere
as the X-ray emission, albeit at a narrower range of altitudes.
Alternatively, any theory wherein GPs are produced in similar parts of
the magnetosphere as ordinary emission must explain their disparate
polarization properties, and how the narrow emission window of \psrc\
is shifted to a phase where X-ray emission peaks.

\section{Conclusions}\label{conclusions}

We have performed baseband searches for microsecond bursts of emission
from sources in the globular cluster M28.  We detected 25 pulses at
20\,cm and 2 pulses at both 2700 and 3500\,MHz, all from \psrc.  The
emission rate of these giant pulses relative to the emission of
ordinary pulses is approximately half that of PSR~B1937+21.  The
pulses occur in two narrow regions of phase that coincide with the
X-ray peaks, but occur on the trailing edges of the radio components.
At 2700\,MHz individual pulses were detected at phases where the
corresponding component of the integrated emission was too weak to
detect.  The integrated radio emission in the main emission regions
has a very steep spectrum of up to $\alpha_{\rm spec} \sim -4.4$.  The
X-rays are emitted over a wider phase range than the giant pulses, so
the two emission phenomena likely originate from similar regions in
the magnetosphere but may not originate from the same physical
mechanism.

The giant pulses have polarimetric properties that are completely
different to those of ordinary emission.  The integrated emission has
no discernable circular polarization, but the giant pulses are up to
60\% circularly polarized.  Their linear polarization seems to have no
preferred orientation, and a variety of position angles are observed.
This could mean that the magnetic field in the emission regions is
variable and overwhelms the dipole field of the pulsar.  Some pulses
are 100\% elliptically polarized; the average polarization fraction
being 79\%.

The two pulses observed at 3500\,MHz have different widths and this
may be interpreted as being due to resolution of the envelopes of
nano-pulses.  Alternatively, the emission volumes may differ.  One of
the pulses is resolved to be 20\,ns wide.  The strongest pulse seen at
2700\,MHz had an inferred brightness temperature at maximum of
$5\times 10^{37}$\,K.  Although individual pulses are seen over
800\,MHz of bandwidth, within 64\,MHz bands at 20\,cm the intrinsic
pulse-intensity fluctuates significantly.  This indicates that the
giant pulses could consist of nano-pulses that are individually narrow
band.

\acknowledgements

Parkes Observatory is funded by the Commonwealth of Australia for
operation as a National Facility managed by CSIRO.  The staff of
Parkes Observatory are thanked for their assistance in helping us
obtain the data used in this paper.  We thank A. Hotan, W. van
Straten, and C. West for observing and computing assistance.  We thank
A. Rots for providing high energy data.  HSK acknowledges the support
of a CSIRO Postgraduate Student Research Scholarship.

\clearpage

\bibliographystyle{apj}

\clearpage

\begin{deluxetable}{llccccccc}
\tabletypesize{\scriptsize}
\tablecolumns{9} 
\tablewidth{0pc} 
\tablecaption{ Search parameters.\label{tab:tsys}} 

\tablehead{ 
& 
& 
\colhead{$\nu$} & 
\colhead{SEFD} & 
\colhead{$T_{1}$} & 
\colhead{$T_{2}$} & 
\colhead{$S_{\rm A}$} & 
\colhead{$E_{\rm lim}$} & 
\colhead{$E_{\rm lim}$} \\[3pt] 
\colhead{Date} & 
\colhead{Receiver} & 
\colhead{(MHz)} & 
\colhead{(Jy)} & 
\colhead{(s)} & 
\colhead{(s)} & 
\colhead{(mJy)} & 
\colhead{(\jyms)} & 
\colhead{($\langle E \rangle _{\rm A}$)} \\[3pt] 
\colhead{(1)} & 
\colhead{(2)} & 
\colhead{(3)} & 
\colhead{(4)} & 
\colhead{(5)} & 
\colhead{(6)} & 
\colhead{(7)} & 
\colhead{(8)} & 
\colhead{(9)} \\ 
}

\startdata 
2003 February 10 & Multibeam                  & 1341      & 39     & 6700   & 320& 2.7    & 91   & 11    \\
       	         &        	              & 1405      & 38     & 6700   & 490& 1.7    & 89   & 18    \\
2003 April    15 & H-OH                       & 1341      & 49     & 7700   &1200& 2.6    & 113  & 14    \\
                 &                            & 1720      & 46     & 7700   &1300& 1.6    & 105  & 22    \\
2003 December 14 &10\,cm feed of 10/50 coaxial& 2700      & 45     & 9200   &730 & 0.20   & 103  & 170   \\
                 &                            & 3500      & 45     & 9200   &760 & 0.12   & 103  & 280   \\
2004 November 11 & Multibeam                  & 1341      & 39     & 2800   & 220& 1.8    & 91   & 17    \\
2004 December 12 & Multibeam                  & 1405      & 38     & 670    & 17 & 4.4    & 89   & 7     \\
\enddata

\end{deluxetable}

\clearpage

\begin{deluxetable}{lccc}
\tabletypesize{\scriptsize}
\tablecolumns{4} 
\tablewidth{0pc} 
\tablecaption{ Fractional Polarizations of Giant Pulses.\label{tab:pol}} 

\tablehead{ 
\colhead{Polarization Type} & 
\colhead{Min} & 
\colhead{Mean} & 
\colhead{Max} \\[3pt] 
\colhead{(1)} & 
\colhead{(2)} & 
\colhead{(3)} & 
\colhead{(4)} 
}

\startdata 
Circular       & 0.0 & 0.27 & 0.6  \\
Linear         & 0.4 & 0.73 & 1 \\
Total          & 0.4 & 0.79 & 1 \\
Ratio of circular to linear& 0.0 & 0.36 & 0.8  \\
\enddata
\end{deluxetable}

\clearpage

\clearpage

\clearpage

\clearpage

\clearpage

\clearpage

\clearpage

\clearpage

\end{document}